\documentclass[english,showpacs,floatfix,11pt]{revtex4}
\usepackage[dvips]{graphicx}
\usepackage{longtable}
\usepackage{amsmath,amssymb}
\usepackage{dcolumn}
\usepackage{latexsym}
\usepackage{babel}
\usepackage[latin1]{inputenc}
\usepackage{color}
\usepackage[colorlinks]{hyperref}

\def\al{\alpha}

\def\kp{\kappa}
\def\nb{\nabla}
\def\pa{\partial}

\def\Ga{\Gamma}
\def\ga{\gamma}
\def\be{\beta}
\def\dl{\delta}
\def\Dl{\Delta}

\def\th{\theta}

\def\Up{\Upsilon}
\def\N{{\cal N}}

\def\diag{\mbox {diag}}

\def\ln{\mbox {ln}}
\def\wt{\widetilde}
\def\l{\left}
\def\r{\right}
\def\H{\textbf{H}}

\def\g{{\sqrt g}}

\begin{document}
\title{\Large Natural Coordinate System in Curved Space-time}
\date{28th October 2017}
\author{Ying-Qiu Gu}
\email{yqgu@fudan.edu.cn} \affiliation{School of Mathematical
Science, Fudan University, Shanghai 200433, China} \pacs{04.20-q,
04.20.Dw, 04.20.Jb, 04.25-g}

\begin{abstract}
In this paper we establish a generally and globally valid coordinate
system in curved space-time with the simultaneous hypersurface
orthogonal to the time coordinate. The time coordinate can be
preseted according to practical evolving process and keep
synchronous with the evolution of the realistic world. In this
coordinate system, it is convenient to express the physical laws and
to calculate physical variables with clear geometrical meaning. We
call it ``natural coordinate system". The constructing method for
the natural coordinate system is concretely provided, and its
physical and geometrical meanings are discussed in detail. In NCS we
make classical approximation of spinor equation to get Newtonian
mechanics, and then make weak field approximation of Einstein's
equation and low speed approximation of particles moving in the
space-time. From the analysis and examples we find it is a nice
coordinate system to describe the realistic curved space-time, and
is helpful to understand the nature of space-time.

\vskip3mm\noindent{Keywords: {\em coordinate system,
orthogonalization, simultaneity, canonical metric}}
\end{abstract}

\maketitle

\section{Introduction}
\setcounter{equation}{0} The selection of coordinate system in
general relativity is important to conveniently express and solve
physical equations. In a suitable coordinate system, the physical
equations have simple and concise forms and definite physical
meanings. In general relativity, the special coordinate systems we
usually used for theoretical discussion are Gaussian normal
coordinate system and harmonic coordinate system\cite{wnbg}.
However, the Gaussian normal coordinate system only exists locally.
The geometrical meaning of harmonic coordinate system is unclear.
Other coordinate systems, such as Weyl-Lewis- Papapetrou
one\cite{rot}, all depend on special structure of space-time.

A more important problem is that, the realistic space-time is an
evolving  Lorentz manifold which has one and only one simultaneous
hypersurface $f(x^\mu)=C$\cite{sbt,prd}. The hypersurface forms the
realistic space and evolves from the present state to the next one.
Since the evolution of the world has probabilistic characteristic of
quantum mechanics, it is obviously non-unique. In principle, it is
most natural and reasonable to describe physical laws in the
coordinate system with simultaneity $t=t_k$ coincident with the
evolving hypersurfaces. Besides, it is perfect if the time
coordinate is orthogonal to this simultaneous hypersurface similarly
to the case in Gaussian normal coordinate system, because only in
such coordinate system, we can conveniently discuss the Hamiltonian
formalism of dynamics and calculate N\"other's charges. We call the
coordinate system with such two features ``natural coordinate system
({\bf NCS})". The purpose of this paper is to prove the existence of
NCS and look for a general method to establish such NCS, and then
clarify its physical meanings.

To get canonical quantification of gravity\cite{adm,hmt} or
numerical simulation of Einstein' field equation\cite{num1,num2}, we
also need to define a simultaneous Cauchy hypersurface. The current
treatment is the ADM decomposition of space-time\cite{adm}. In ADM
decomposition, the metric is reduced into time plus space with a
shift vector $N^j$,
\begin{eqnarray}
ds^2=\N^2 dt^2 +\wt g_{kl}(dx^k+N^k
dt)(dx^l+N^ldt),\label{crtr}\end{eqnarray}where the repeated Latin
indices means summation of spatial indices. The Ricci tensor and
scalar curvature can be expressed in simple and reduced from. If we
can set the shift vector $N^j= 0$ and clarify the specific
geometrical  meaning of space and time, the formalism will be even
simpler.

\section{Construction of Natural coordinate system}
\setcounter{equation}{0}

At first, we discuss the problem of orthogonalization. We use the
Greek characters such as $\mu,\nu$ to stand for the 4-d coordinate
indices, and Latin characters such as  $k,l$ for 3-d indices. Assume
the metric of the space-time is given by
\begin{eqnarray}
ds^2=\wt g_{tt}dt^2 +2A_kdt dx^k + \wt g_{kl}dx^k dx^l,\quad
A_k\equiv \wt g_{0k},\label{crtr}\end{eqnarray} we have the
following conclusion for  orthogonalization.

{\bf Theorem 1.} {\em Keeping the time coordinate $t$ not
transformed, for (\ref{crtr}) there exists a class of regular
spatial coordinate transformation
\begin{eqnarray}
x^k=x^k(t,y^l),\quad dx^k = \frac {\pa x^k}{\pa t} dt +\frac {\pa
x^k}{\pa y^l} dy^l,\label{xtoy}\end{eqnarray} such that `$dt$' is
orthogonal to `$ dy^k $', i.e., in the new coordinate system we
have}
\begin{eqnarray}
ds^2=g_{tt}dt^2- g_{kl}dy^k dy^l.\label{ntr}\end{eqnarray}

{\bf Proof.} Substituting (\ref{xtoy}) into (\ref{crtr}), we get
\begin{eqnarray}
ds^2= g_{tt}dt^2 +2B_kdt dy^k - g_{kl}dy^k
dy^l.\label{crn0}\end{eqnarray} in which
\begin{eqnarray}
B_k\equiv \l(\wt g_{nl} \frac {\pa x^l}{\pa t}+A_n\r)\frac {\pa
x^n}{\pa y^k}.\label{Ak}\end{eqnarray} Since the transformation
(\ref{xtoy}) should be invertible,  the Jacobian matrix $\frac {\pa
x^l}{\pa y^k}$ also is invertible. Then $B_k=0$ is equivalent to
\begin{eqnarray}
{\pa_t x^k}=-\bar g^{kl} A_l=-\bar g^{kl} \wt
g_{0l},\label{cdtn}\end{eqnarray} where $\bar g^{kl}$ is the
inversion of spatial metric, i.e., $\wt g_{kl}\bar g^{ln}=\dl^n_k$.

Since the time coordinate $t$ is a fixed independent coordinate,
(\ref{cdtn}) becomes a first order ordinary differential equation
system of $x^k$. For the initial value problem we have unique
solution
\begin{eqnarray}
x^k=f^k(t,X^l),\label{trns0}\end{eqnarray} in which $X^k$ is
integral constants determined by initial values. Making an arbitrary
invertible and differentiable coordinate transformation
\begin{eqnarray}
X^k=F^k(y^l),\label{trns1}\end{eqnarray} and substituting it into
(\ref{trns0}) we get (\ref{xtoy}). The proof is finished.

Obviously, if $g_{kl}$ in (\ref{ntr}) is independent of $t$, the
above procedure of orthogonalization can proceed, such that  $dy^1$
is orthogonal to the other two coordinates. Since the proof  is
constructive, it can also be used as method to establish the NCS.

In what follows we examine the problem of simultaneity. In
\cite{sbt,prd}, we have proved that, in the realistic space-time
there exists one and only one simultaneous hypersurface
$f(x^\mu)=C$, which is an objective existence independent of
coordinate system. According to its physical meanings, the
hypersurface should have the following two features: F1. Assuming
$dx^0$ is time-like, by the unidirection of time we have $\pa_0f>0$
under a suitable function transformation of $f(x^\mu)$. F2.
$f(x^\mu)$ has higher smoothness than $C^1$, this is because all
physical dynamics consists of first or second order differential
equation system.  We take F1 and F2 as basic assumptions for the
simultaneous hypersurface. Redefine the time coordinate
\begin{eqnarray}
t=f(x^\mu),\label{time}\end{eqnarray} then the simultaneity of the
real world becomes $t\equiv t_k$, which defines the global realistic
simultaneity. We call it the ``{\bf cosmic time}"\cite{quat}.
Clearly, if we make orthogonalization based on this cosmic time, we
get the NCS (\ref{ntr}) with special significances. We call the
space orthogonal to the cosmic time as ``{\bf cosmic space}".
Obviously, the coordinate system for cosmic time and cosmic space
can be determined to the following arbitrary regular transformation
\begin{eqnarray}
t'=T(t),(\pa_t T>0),\qquad y'^k=Y^k(y^l).\label{und}\end{eqnarray}
However, the time element $d\tau =\sqrt{g_{00}} dt$ and the spatial
volume element $dV=\sqrt{\bar g}d^3y$ are objective quantities
independent of transformation (\ref{und}), where $\bar
g=\det(g_{kl})$. Then we proved,

{\bf Theorem 2.} {\em In the realistic world we have a unique global
cosmic time $t$ and a unique cosmic space orthogonal to this time.
The objective time is calculated by $d\tau =\sqrt{g_{00}} dt$, and
the objective volume of the space is calculated by  $dV=\sqrt{\bar
g}d^3y$. They are independent of coordinate system.}

In contrast with AMD decomposition, in NCS the shift vector is
removed and the time and space have endowed with concrete physical
meanings.  The NCS (\ref{ntr}) looks like Gaussian normal coordinate
system, but their physical meanings are different: M1. NCS is always
globally valid, and $g_{tt}$ represents gravity which cannot be
merged into coordinate $t$ in the case $\pa_k g_{tt}\not\equiv 0$.
M2. The construction is different. NCS is independent of geodesic of
concrete particles. M3. The time $t$ in NCS corresponds to realistic
cosmic time, which forms a global standard of time for all
particles. Since the construction of NCS does not need any special
conditions, we can directly express physical laws in this coordinate
system without any declaration.

Now we concretely construct a NCS from a non-diagonal Kerr-type
metric\cite{rot}. More generally, the metric takes the following
form in the coordinate system $(t, r, \th, \phi)$,
\begin{equation} \wt g_{\mu\nu}=\left( \begin {array}{cccc} {u}^{2}&0&0&uw\\ \noalign{\medskip}0&-a&0&0\\
\noalign{\medskip}0&0&-b&0\\\noalign{\medskip}uw&0&0&w^2-v
\end {array} \right),\qquad \wt g^{\mu\nu}= \left( \begin {array}{cccc} {\frac {-{w}^{2}+v}{{u}^{2}v}}&0&0&{
\frac {w}{uv}}\\ \noalign{\medskip}0&-{a}^{-1}&0&0
\\ \noalign{\medskip}0&0&-{b}^{-1}&0\\ \noalign{\medskip}{\frac {w}{uv
}}&0&0&-{v}^{-1}\end {array} \right). \label{3.10} \end{equation}
where $u,v,w,a,b$ are smooth functions of $(r,\th)$, but independent
of $(t,\phi)$. However we should pay attention to that
\begin{equation} \wt g_{kl}=\left( \begin {array}{cccc} -a&0&0\\
\noalign{\medskip}0&-b&0\\\noalign{\medskip}0&0&w^2-v
\end {array} \right),\qquad \bar g^{kl}= \left( \begin {array}{ccc} -{a}^{-1}&0&0
\\ \noalign{\medskip} 0&-{b}^{-1}&0\\ \noalign{\medskip} 0&0&-{(v-w^2)}^{-1}\end {array} \right). \label{3.10*} \end{equation}
Assuming that $t$ is the cosmic time, then by (\ref{cdtn}) we have
\begin{eqnarray}
\pa_t r=0,\quad \pa_t \th =0, \quad\pa_t \phi =- \bar
g^{\phi\phi}\wt g_{t\phi}={(v-w^2)}^{-1} u
w.\label{time}\end{eqnarray} The simplest solution is given by
\begin{eqnarray}
r=r',\quad  \th =\th', \quad \phi ={(v-w^2)}^{-1} u w
t+\phi'.\label{time}\end{eqnarray} It is easy to check, the metric
(\ref{3.10}) is converted into the standard one (\ref{ntr}) in new
coordinate system $(t,r',\th',\phi')$. The coordinate $\phi'$
becomes an evolving one. The clock keeping static to this coordinate
system goes in uniform speed. This is somewhat similar to the Mach's
principle\cite{mach1,mach2}.

\section{applications}
\setcounter{equation}{0} \subsection{Classical Approximation of
Spinor Equation}

To discuss the Hamiltonian formalism of Dirac equation in curved
space-time we need the cosmic time coordinate with clear physical
meaning and orthogonal to space. To calculate the N\"other charges
of a spinor such as energy-momentum and velocity, we need to do
integration on the realistic cosmic space. Some works have been done
in Gaussian normal coordinate system\cite{spn}. Now we define some
classical concepts in NCS to show the advantages.

Assume the element of the space-time satisfies
\begin{eqnarray}
d\mathbf{x}=\wt\ga_\mu dx^\mu=\ga_a \dl X^a,\quad \wt \ga_\mu
=l_{\mu}^{~a}\ga_a,\quad \wt \ga^\mu =h^{\mu}_{~a}\ga^a.
\label{1.1a}\end{eqnarray} in which the tetrad  $\ga_\al$ and
$\wt\ga_\mu$ satisfy the following $C\ell_{1,3}$ Clifford algebra,
\begin{eqnarray}
\ga_a\ga_b+\ga_b\ga_a=2\eta_{ab},\qquad
\wt\ga_\mu\wt\ga_\nu+\wt\ga_\nu\wt\ga_\mu=2g_{\mu\nu}. \label{1.2a}
\end{eqnarray}
In NCS with metric $\diag(g_{00},-\bar g_{jk})$,  we have
\begin{eqnarray}
l^0_{~0}=\sqrt{g_{00}},\quad h^0_{~0}= \sqrt{g^{00}}, \quad
h^0_{~k}= l_0^{~k}=0.
\end{eqnarray} Then we get spinor connection as\cite{spn}
\begin{eqnarray}
\Up_\mu&=& \frac 1 2 \l(l_0^{~0} \pa_t h^0_{~0}+\pa_t\ln\sqrt{g},
~\vec l_k \cdot \pa_j\vec h^j+\pa_k\ln\sqrt{g}\r). \label{4.2}
\end{eqnarray}
In NCS, to lift and lower the index of a vector means $\Up^0=g^{00}
\Up_0, \Up^k=-\bar g^{kl}\Up_l$.

We consider Dirac equation with electromagnetic potential $eA^\mu$.
Its Hamiltonian formalism is given by
\begin{eqnarray} \al^0 i(\pa_t+\Up_t)\phi=\H\phi, \label{4.3*}
\end{eqnarray}
where the Hamiltonian is defined by
\begin{eqnarray}
\H= -\al^k \cdot[ i(\pa_k+\Up_k)-e A_k]+e\al^0 A_0+m \ga_0,\quad
(\al^\mu\equiv \ga_0 \wt\ga^\mu).
\end{eqnarray}

Similarly to the case in flat space-time, we define the coordinate
$X^k$ and speed $v^k$ of the spinor as follows\cite{spn,eng},
\begin{eqnarray}
X^k(t)\equiv \int_{S^3}  x^k q^0 \sqrt{g}d^3x,\quad  v^k\equiv \frac
d {dt} X^k,\label{4.4}
\end{eqnarray}
where $S^3$ stands for the total simultaneous hypersurface, i.e.,
the cosmic space, $q^\mu$ is the current vector $q^\mu=\phi^+\al^\mu
\phi$. By the definition (\ref{4.4}) and the current conservation
law $q^\mu_{;\mu}=0$, it is easy to check
\begin{eqnarray}
 v^k = \int_{S^3}  x^k \pa_t(q^0 \sqrt g)  d^3x=\int_{S^3}
x^k q^0_{;t} \sqrt g  d^3x=-\int_{S^3}  x^k q^l_{;l} \sqrt g d^3x =
\int_{S^3} q^k\sqrt{g} d^3x. \label{4.6}
\end{eqnarray}
With the normalizing condition $\int_{S^3}q^0\sqrt{g} d^3x=1$, we
have  the point-particle model,
\begin{eqnarray}
q^\mu\to u^\mu \sqrt{g_{00}-\bar g_{kl}v^k v^l}\dl^3(\vec x-\vec
X),\qquad u^\mu\equiv \frac {d X^\mu}{d\tau}=(1, \vec v)/
\sqrt{g_{00}-\bar g_{kl}v^k v^l}, \label{4.6*}
\end{eqnarray}
where the Dirac-$\dl$ means $\int_{S^3}\dl^3(\vec x-\vec
X)\sqrt{\bar g} d^3x=1$ and  $\tau$ is the proper time of the
particle $d\tau = \sqrt{g_{00}-\bar g_{kl}v^k v^l } dt$.

Clearly, the particle static to NCS has largest proper time, i.e.,
its time goes fastest. Since the time of any particle should go at a
finite speed, which certainly have a least upper bound everywhere,
this requirement also discloses the existence of a special
coordinate system in Nature. By the definition of NCS, we find that
the cosmic time and cosmic space are quite near the concepts of
Galilean and Newtonian absolute space-time. The classical concept of
space-time cannot be completely cast away.

Define the 4-dimensional momentum of the spinor by \begin{eqnarray}
p^\mu\equiv \Re \int_{S^3}\phi^+\al^0\hat p^\mu \phi\sqrt{ g}
d^3x=\Re \int_{S^3}\phi^+\hat p^\mu \phi\sqrt{\bar g} d^3x,\quad
\hat p^\mu\equiv i(\pa^\mu+\Up^\mu)-eA^\mu. \label{4.10}
 \end{eqnarray}
For a spinor at energy eigenstate, we have the classical
approximation $p^\mu = m u^\mu$, where $m$ defines the classical
inertial mass of the spinor.  The classical approximation of
energy-momentum tensor $T^{\mu\nu}$ for a free spinor is given by
\begin{eqnarray}
T^{\mu\nu} \to( m u^\mu u^\nu+w g^{\mu\nu})\sqrt{1-g^{00}\bar
g_{kl}v^k v^l }\dl^3(\vec x-\vec X),\label{clst}
\end{eqnarray}
where $w$ is a constant to representing self-potential, and $w=0$
corresponds to linear spinor. By $T^{\mu\nu}_{~;\nu}=0$ we get
\begin{eqnarray}
\pa_\nu(T^{\mu\nu}\sqrt{ g})+\Ga^\mu_{\al\be}T^{\al\be}{\sqrt{
g}}=0.\label{csv0}
\end{eqnarray}
The  integral form is given by
\begin{eqnarray}
\frac d {dt} \int_{S^3} T^{\mu 0}\sqrt{g} d^3 x+ \int _{S^3}
\Ga^\mu_{\al\be}T^{\al\be}\sqrt{ g} d^3 x=0.\label{csv1}
\end{eqnarray}
Substituting (\ref{clst}) into (\ref{csv1}) and noticing
${g}={g_{00}\bar g}$, we get the Newton's second law for a free
spinor
\begin{eqnarray}
\frac d {dt} \l(m u^\mu +wg^{\mu 0} \sqrt{g_{00}-\bar g_{kl}v^k v^l
}\r)+\Ga^\mu_{\al\be}\l(m u^\al u^\be+w
g^{\al\be}\r)\sqrt{g_{00}-\bar g_{kl}v^k v^l }=0. \label{gds0}
\end{eqnarray}
In the case of linear spinor we have $w=0$, and then (\ref{gds0})
becomes the geodesic equation.  From the above calculation we find
that, some N\"other's charges of fields can be clearly defined and
computed only in NCS. So NCS set up a natural connection between
quantum mechanics and classical one.

\subsection{Linearization of  Einstein's Field Equation}
In the case of weak gravity and low speeds of particles,  we make
linearization of Einstein's field equation and get a set of wave
equation in  harmonic coordinate system with  the de Donder
coordinate condition $\Ga^\mu=0$, where
\begin{eqnarray}
\Ga^\mu \equiv g^{\al\be}\Ga^\mu_{\al\be}=-\frac 1 \g \pa_\nu(\g
g^{\mu\nu}). \label{a1.16}
\end{eqnarray}
In this case the metric is near the Minkowski metric
\begin{eqnarray}
\eta_{\mu\nu}=\eta^{\mu\nu}=\diag(1, -1, -1, -1). \label{a2.1}
\end{eqnarray}
For weak-field approximation, we have the linearization for the
metric\cite{wnbg}
\begin{eqnarray}
g_{\mu\nu}&\equiv &\eta_{\mu\nu}+h_{\mu\nu},\quad
g^{\mu\nu}\dot =\eta^{\mu\nu}-h^{\mu\nu},\label{a2.3}\\
h^{\mu\nu}&=&\eta^{\mu\al}\eta^{\nu\be}h_{\al\be},\quad~
h=h^\mu_{~\mu}=\eta^{\mu\nu}h_{\mu\nu},\label{a2.2}\\
g &\dot =& 1+h,\qquad\quad  \g \dot = 1+\frac 1 2 h.\label{a2.4}
\end{eqnarray}
In what follows, we directly use $=$ to replace $\dot =$ for
convenience. By straightforward calculation, we get the
linearization for other parameters
\begin{eqnarray}
\Ga^\mu_{\al\be} &=& \frac 1 2\eta^{\mu\nu}(\pa_\al h_{\nu\be}+\pa_\be h_{\al\nu}-\pa_\nu h_{\al\be}),\label{a2.5}\\
\Ga^\mu&=& \pa_\nu \chi^{\mu\nu},\qquad \chi^{\mu\nu}\equiv h^{\mu\nu}-\frac 1 2 \eta^{\mu\nu}h, \label{a2.6} \\
R_{\mu\nu}&=&\frac 1 2 \pa_\al\pa^\al h_{\mu\nu}-\frac 1 2
(\eta_{\mu\al}\pa _\nu \Ga^\al
+ \eta_{\nu\al}\pa _\mu \Ga^\al) \label{a2.7},\\
R&=&\frac 1 2 \pa_\al\pa^\al h -\pa _\al \Ga^\al, \label{a2.9}\\
G^{\mu\nu} &=& \frac 1 2\l( \pa_\al \pa^\al \chi^{\mu\nu} -
\pa^\mu\Ga^\nu- \pa^\nu\Ga^{\mu}+\eta^{\mu\nu}\pa_\al\Ga^\al
\r),\label{a2.12}
\end{eqnarray}
in which $\pa_\al\pa^\al=\pa_t^2-\Dl$ is the d'Alembert operator. By
(\ref{a2.12}) we find the Bianchi identity $\pa_\mu G^{\mu\nu}=0$
holds.

In NCS the time coordinate is fixed, so we have not coordinate
condition $\Ga^\mu=0$ in general. However, by $G^{0\mu}$ we find
$\Ga^\mu$ can be easily solved independently in NCS. Explicitly we
have
\begin{eqnarray}
\Ga^0 &=& \pa_t \Psi, \qquad  \Ga^k = \frac 1 2 \pa_k h +\pa_n
h_{kn},\label{gak}
\\ \Psi &\equiv& \chi^{00}= \frac1 2
(h_{00}+h_{11}+h_{22}+h_{33}), \label{dff}
\end{eqnarray}
where the repeated Latin indices means summation of spatial indices.
Then by $G^{0\mu}$ we get independent equations for $(\Psi, \Ga^k)$
as follows
\begin{eqnarray}
G^{00}&=& -\frac 1 2 \Dl \Psi=\kp T^{00},\\
G^{0k}&=&-\frac 1 2 \pa_t (\Ga^k-\pa_k\Psi)=\kp T^{0k}.
\label{a2.11}
\end{eqnarray}
By the above equations we can determine the intermediate variables
$\Ga^\mu$. Substituting it into (\ref{a2.12}) we get usual wave
equations for $h_{kl}$.

From the procedure we find the coordinate condition is determined by
dynamics as well as the initial and boundary values. This is
natural, because the simultaneous hypersurface is evolving itself.

Now we examine the classical particles moving in the weak gravity.
Under low speed assumption we omit $O(v^2)$. Noticing $v^k\equiv
\frac d {dt}X^k=\sqrt{1+h_{00}} \frac d {d\tau} X^k$, the geodesic
equation becomes
\begin{eqnarray}
\frac d {dt} v^k = - \pa_k \Phi +v^k\pa_t\Phi +v^n \pa_t
h_{kn},\label{gsc1}
\end{eqnarray}
where $\Phi =\frac 1 2 h_{00}$ is the Newtonian gravitational
potential. For many such particles move without collision similarly
to stars in a galaxy, which form  zero-pressure and inviscid fluid
with following energy-momentum tensor
\begin{eqnarray}
T^{\mu\nu}=\rho U^\mu U^\nu,\label{a1.2}
\end{eqnarray}
where $\rho$ is the comoving mass density of the stars, and $U^\mu$
is the 4-vector speed of the stellar flow. We have energy-momentum
conservation law $T^{\mu\nu}_{;\nu}=0$. Expressing it in the form of
equations of continuity and motion, we get the dynamical equations
for the stars
\begin{eqnarray}
U^\mu \pa_\mu \rho + \rho_sU^\mu_{~;\mu}=0,\qquad U^{\nu}
U^\mu_{~;\nu}=0. \label{a1.3}
\end{eqnarray}
Define the stellar speed $\vec V$ by
\begin{eqnarray}
\vec V\equiv \frac 1{U^0}(U^1,U^2,U^3), \label{a2.16}
\end{eqnarray}
which is approximately equivalent to the usual definition of
velocity $\frac d{dt}\vec X$. By line element equation we have
\begin{eqnarray} 1=\sqrt{g_{\mu\nu} U^\mu U^\nu}= ( 1+2 \Phi +g_{kl} V^k V^l )^{\frac 1 2} U^0.\label{a2.17}
\end{eqnarray}
Omitting $O(V^2)$ terms we get the low-speed assumption
\begin{eqnarray} U^0 = 1 - \Phi.\label{a2.18}
\end{eqnarray}
Substituting (\ref{a2.16}) and (\ref{a2.18}) into (\ref{a1.3}) and
omitting the high order terms, we get the continuity equation and
motion equation for stars
\begin{eqnarray}
 \frac d{dt}\rho &=& - \rho \nb\cdot \vec
V-\rho
(\pa_t+\vec V\cdot \nb)\Psi, \label{lxfc}\\
\frac d{dt} V^k &=& - \pa_k \Phi+  V^k\pa_t
\Phi+V^n\pa_th_{kn},\label{ndfc}
\end{eqnarray}
where $\frac d {dt} =\pa_t+\vec V\cdot \nb$. (\ref{ndfc}) and
(\ref{gsc1}) have the same form. In NCS, the gravitomagnetic effect
vanishes. However, this effect exists in other coordinate system.
There are many researches on this
problem\cite{grm1,grm2,grm3,grm4,grm5,glx}. In usual coordinate
system (\ref{crtr}), we have gravitomagnetic force $\vec V\times
(\nb\times\vec A)$ similar to Lorentz force in electromagnetism.

\section{discussion and conclusion}
\setcounter{equation}{0}

In this paper, we establish a generally and globally valid natural
coordinate system in curved space-time. In this coordinate system,
the time and space keep synchronous with the evolution of the
realistic world, which have special physical and philosophical
meanings.  In NCS we can discuss Hamiltonian formalism conveniently
as done under ADM decomposition. As examples, we simply make
classical approximation of spinor equation to get Newtonian
mechanics, which clearly shows the relationship between quantum
mechanics, classical mechanics and general relativity. By weak field
approximation of Einstein's equation and low speed approximation of
particles moving in the space-time, we find coordinate condition is
included in dynamical equations, and the motion equation of
particles has a simple form, in which the gravitomagnetic force
vanishes. From the analysis and examples we find the NCS is a nice
coordinate system to describe the realistic curved space-time, and
is helpful to understand the nature of space-time.

\section*{Acknowledgments}
The author is grateful to his supervisor Prof. Ta-Tsien Li and Prof.
Han-Ji Shang for their encouragement.

\end{document}